\documentstyle[11pt]{article}
\begin{document}

\title{{\bf Adsorption of line segments on a square
lattice}} \author{{\bf B. Bonnier, M. Hontebeyrie,
Y. Leroyer,} \\
 {\bf C. Meyers and E. Pommiers}\\[1cm]
{\em Laboratoire de Physique
Th\'eorique} \\ {\em CNRS, Unit\'e Associ\'ee 764}\\
{\em Universit\'e de Bordeaux I}\\
{\em 19 rue du Solarium, F-33175 Gradignan Cedex}}
\date{}
\begin{titlepage}
\maketitle
\thispagestyle{empty}

\begin{abstract}
We study the deposition of line segments on a two-dimensional square
lattice. The estimates for the coverage at jamming  obtained
by Monte-Carlo simulations and by $7^{th}$-order time-series
expansion are successfully compared.
The non-trivial limit of adsorption of infinitely long segments  is
studied, and the lattice coverage is consistently obtained  using
these two approaches.
 \end{abstract}
\vfill
LPTB 93-8\\
June 1993\\
PACS 05.70Ln, 68.10Jy\\
E-Mail:
{\bf Leroyer @FRCPN11.IN2P3.FR}
\end{titlepage}

\section{Introduction}
Random sequential adsorption (RSA) has been used for a long time as a
model of irreversible deposition processes~\cite{Evans,Privman}. An
object of a given shape is placed
randomly on a substrate, subject to the constraint that it does not
overlap previously
deposited objects. One determines the coverage, defined as the fraction
of area
covered by the adsorbed objects, as a function of time, and its
infinite time limit, called the jamming limit. These quantities can be
determined exactly in one dimension and by approximate methods or
numerical simulations in higher dimensions. The RSA models are
characterised by
two main parameters~: the nature of the substrate - discrete or
continuous - and
the shape of the deposited objects. Although lattice models are less
directly connected with experimental situations, they are easily
accessible to numerical simulations and accurate results can be
obtained from them. Furthermore,
in most cases, one can define a {\em scaling limit} which allows an
extrapolation to the continuum. The jamming limit for the deposition
of $k$-mers on
a one dimensional lattice has been known exactly for a long
time~\cite{Flory} as well
as its continuum limit~\cite{Renyi}. In two dimensions, the regular
objects that fit a square lattice are rectangles of width $w$ and
length
$\ell$ expressed in units of the lattice spacing. When the aspect
ratio $\alpha$, defined as the length-to-width ratio, varies from 1
to
$\infty$ the shape of the objects changes from the square to
the line segment. The continuum limit of aligned squares, taken
by letting the edge size approach infinity with the aspect
ratio fixed to one, has been studied by Privman, Wang and
Nielaba\cite{PWN} and by Brosilow et al~\cite{Brosilow}
in the framework of an extensive numerical simulation, leading
to
an accurate determination of the saturation coverage and
of the correlation functions. On the other hand, the deposition
of randomly oriented rectangles in the continuum  has received
much attention in order to guess the
influence of the aspect ratio on the kinetics of the process
and on its
jamming limit~\cite{Sherwood,Vigil,Schaaf}. For instance it has
been shown
 that in the case of
extremely elongated objects, the jamming coverage approaches zero
as a power of the
inverse of the aspect ratio. In the limit of infinite aspect ratio
which corresponds to the deposition of {\em randomly oriented}
line segments, the time evolution of the density of deposited
segments       is driven by a power-law
  behaviour~\cite{Sherwood,Vigil2,Tarjus,Ricci}.

RSA of dimers or of small $k$-mers on the lattice, being one of
the
simplest two-dimensional process, has been widely studied as a testing
model for the approximation methods and numerical simulations. In
contrast to the continuum case, the RSA of long segments on a
bidimensional lattice has not yet been
systematically investigated~\cite{Mana} and this is the problem we
address in this paper.	The study of such systems can give insight
into
experimental situations involving the deposition of rectangle-like
        objects
with
large aspect ratio such like rigid (co-)polymers on a latticized
substrate. In addition, in the limit of infinite segment length, this
model is equivalent to the continuum deposition of {\em aligned} unit
segments onto the plane, which is to be compared to the case of {\em
randomly oriented} ones.

We first attack the problem by means of an extensive numerical
simulation which is presented in section 2 of this paper, where we
give the
technical details and
 the analysis of finite size effects and propose an expression for the
large $k$ behaviour of the jamming limit.
Then in section 3, we derive $7^{th}$ order time-series expansion
 of the coverage analitycally in $k$ and compare the
result of the extrapolation at jamming with the
asymptotic value obtained by the
simulation. We comment our results
in the last section.

\section{The numerical simulation}
\subsection{The method}
We consider a periodic square lattice of linear size $L$, on
which we randomly deposit line segments of $k$ sites ($k$-mers).
Since we are interested in the limit of long segments, we
need a large scale simulation. However, the standard method
usually implemented in the continuum deposition of
oriented squares~\cite{Brosilow}, of dividing the
lattice into cells containing at most one object,
is not efficient here.
Therefore, due to memory storage and computing time
limitations, we had to restrict our simulation to
segments of length $k\leq 512$ sites on lattices of
linear size $L\leq
4096$, preserving in all cases a ratio $L/k\geq 8$. A
subsequent
study of finite size effects allows us to make a reliable
extrapolation to the $k\rightarrow \infty$ limit. This point will
be discussed
below.

The time evolution to the jamming limit is divided in two
regimes›cite{Privman} : in the initial stage the possible
adsorption site is randomly chosen among {\em all}
the sites; in the late stage, it is drawn from
the list of vacant positions, which is regularly updated after a
fixed number of depositions. The relative duration of both stages is
optimized in such a way that the rejection rate in the first stage
remains low,
whereas, in the second stage, the list of vacant
positions is small enough for the updating procedure
to be short. Good
balance between these constraints is realised for a number of
attempts during the first stage of typically four to five times the
total number of
sites.

For the largest lattice sizes, the memory storage needed is
very large. We use multispin coding to store the occupancy state of
each site, and initialize the list of vacant positions only at the
beginning of the second stage when it is reduced to less
than five percent of the total number of sites.

Finally, the sample used for averaging and for the error
analysis consists of at least 100 independent runs.
Although rather small, this sample size leads to results
of sufficient accuracy for our purpose.

\subsection{Finite size effects}
 Finite size effects are exactly known in one
dimension~\cite{MacKenzie,Bartelt}.  MacKenzie has shown that, for RSA of
$k$-mers at jamming, the number
of vacant sites on a lattice of $L$ sites with open boundary conditions is
given by
$$ V_L(k)=(k+L)[\bar{V}(k)+O(\frac{1}{(L/k)!})] \qquad {\rm for}\;
L\rightarrow\infty\quad{\rm and}\;
k/L\rightarrow 0$$
where $\bar{V}(k)$ does not depend on $L$. A {\em periodic} lattice of $L$
sites,
once the first segment is deposited,
becomes an open one of $L-k$ sites, and
the coverages for both systems are related by (where the superscript
$(O)$ stands for ``open" and $(P)$ for periodic)~:
 $$\theta^{(P)}_L(k)=1-\frac{V^{(P)}_L(k)}{L} = 1-\frac{V^{(O)}_{L-
k}(k)}{L} = \bar{V}(k)+O(\frac{1}{(L/k)!})$$
It follows that finite-size corrections are less than exponentially
small
 for a periodic lattice.
Although this argument cannot be directly extended to higher
dimensions
we expect that in two dimensions, the edge effects
on the jamming limit on a {\em periodic} lattice decrease very rapidly
when $L\rightarrow\infty$ and $k/L\rightarrow 0$.

Actually, Brosilow et al~\cite{Brosilow} and Privman et
al~\cite{Privman} have already	observed that
 finite-size effects are negligible for the deposition of oriented
squares on periodic lattices of size as small as 8 times the square
edge. In our case, the deposited objects being very asymmetric, we
may expect a stronger dependence on the size of the system. In
addition, in order to reach the large segment limit at lowest
computing cost, we consider systems whose size is not very large
compared to the size of the adsorbed segments. For these reasons we
must handle carefully the finite-size effects. To estimate them,  we
perform the following analysis. For
a given lattice size $L$, we measure
$\theta_L(k)$ for various values of $k$ up to $k\simeq L/2$, and
then repeat this measure for a larger value of $L$. Figure 1 shows
the results for
$L=128$ and $L=256$. It is apparent that both lattices give the same
result for $k\leq 42$. The finite-size effects then start to occur
for the $L=128$ lattice,
whereas they appear only for $k\geq 90$ for the largest lattice. The
continuous line results from the extrapolation discussed in the next
section. It thus appears that a lattice of a given size $L$ behaves as
if it were
infinite as long as
the segment size does not exceed at most a quarter of the lattice
edge. With this rough analysis as a guide,
 we have measured $\theta_L(k)$,  using
several lattice sizes (typically 3 to 4 values) for each $k$ value, up
to $L/k = 4$ and check that
for the largest sizes, the measured coverages remain consistently
within
the error bars. We give the results in Table I, which will be used as
the basis of our extrapolation of the next section.

Another issue concerning the finiteness of the system is  the
standard deviation
over the sample of the fraction of occupied sites, defined by
$$\sigma_\theta=\sqrt{<\theta^2>-<\theta >^2}$$
and connected to the statistical error of the jamming coverage,
$\Delta\theta~=~\sigma_\theta~/\sqrt{N_s}$,  where $N_s$ is the sample
size. On the basis of
standard statistical arguments, if one assume that the fluctuations of
$\theta$ are driven by the variations of the number of deposited
objects, one expect $\sigma_\theta$ to decrease as the inverse
square root of this number, which would give a behaviour
in $\sqrt{k}/L$.
Instead, for the whole set of data,  we observe that
$$\sigma_\theta (k,L)\simeq 0.1\; \frac{k}{L}$$
This behaviour emerges clearly from Table II, where $\sigma_\theta$
measured for several fixed $k/L$, is roughly independant of the
lattice size $L$, the constant
value obtained for each $k/L$ being proportional to $k/L$. It is
confirmed
in Figure 2 where the $k$ dependence of $\sigma_\theta$  for $L=256$
appears to be linear. On the same plot, we have superimposed some data
for $L=512$ and
 $k=16,32,64,128$ which coincide
within the error bars with the $L=256$ data for $k=8,16,32,64$
respectively. The $L$ dependence of $\sigma_\theta$ is a common
feature of all the RSA simulations~\cite{Evans},
whereas a linear $k$ dependence is somewhat unexpected. This result
can
be interpreted as an
indication that in the limit of long segments, the system behaves
mainly as a one-dimensional one along each direction, the fluctuations
being linked to the number of deposited objects {\em along each line},
$(L/k)^2$.

\subsection{The extrapolation}
Following the finite-size effect analysis of the preceding section, we
consider that for each segment length $k$, the value of $\theta_L(k)$
measured on the largest lattice size $L$ of Table I,
is a good  estimate of $\theta_\infty (k)  =
\lim_{L\rightarrow\infty}\theta_L(k)$. These
data, plotted as a function of $k$ are displayed in figure 3 for
$k\geq 24$. In order to extrapolate to $k\rightarrow\infty$, we must
guess the large $k$ behaviour.
Analogous analyses have been performed previously in one
dimension~\cite{MacKenzie}
and in two dimensions for the deposition of
squares~\cite{Privman,Brosilow}. In both cases
the large $k$ limit corresponds to the continuum limit and is
approached up to a  $1/k$ corrective term.

Although our data exclude a single $1/k$ dependence, they are quite
compatible
with a superposition of $1/k$ and $1/k^2$ corrective terms. The best
fit according
to this behaviour, realised for $k\geq 48$, is displayed in Figure 3
and corresponds to the expression~:
\begin{equation}
  \theta_\infty (k) = 0.660+1.071\;\frac{1}{k}-3.47\;\frac{1}{k^2}
\label{MCfit}
\end{equation}
from which we conclude that
\begin{equation}
  \theta_\infty (\infty ) =0.660\pm 0.002      \label{MC}
\end{equation}
where the error bar results from a variation of the fitting interval.
Accordingly, the coefficents of $1/k$ and $1/k^2$ in eq.(\ref{MCfit})
vary in the ranges $[0.85, 1.08]$ and $[-1.0,-3.7]$ respectively.

Let us finally discuss how the jamming coverage varies with the
deposition mechanism. Actually, besides the conventional deposition
mechanism used here, one can choose the so-called ``end-on" mechanism
in which, once a vacant site has been
found, the deposition is (randomly) attempted in {\em all the
directions} untill the segment is adsorbed or rejected. This
method leads to denser configurations than the conventional
one~\cite{Nord} for small $k$-mers, and for infinitely long $k$-mers
we quote here the coverage from  Manna and \u{S}vraki\'{c}~\cite{Mana}
$$\theta_\infty (k) = 0.583(\pm 0.010)+0.32\;\frac{1}{\ln k} \quad ,$$
clearly smaller than our result, Eq.(\ref{MC}).
Comparing both types of data, we observe
 a cross-over for $k\simeq 16$, our saturation coverage becoming
      larger than
the end-on one above this value.
 Let us point out that the same kind of cross-over between
conventional and end-on coverages has recently been observed in one
dimension \cite{NordFG} at $k=4$.
 \section{Perturbative $k$-mer filling of the square lattice}
 \subsection{The perturbative expansion }
{}~
We will   determine the time-series expansion of the coverage
$\theta(k,t)$ for
the  RSA of $k$-mers on a two-dimensional square lattice.
The segment orientation is chosen at random with equal probability
for horizontal and vertical deposition.

We first construct an operator which realizes the sequential addition
process of arbitrary objects on a lattice.
This elementary time-evolution operator is obtained by a
generalization of the quantum mechanical methods used by Fan and
Percus \cite{FP}, and by Dickman, Wang and Jensen \cite{DWJ} to the
deposition of arbitrary objects , and cannot be evaluated except in
a perturbative way.

The perturbative expansion (PE) of $\theta(k,t)$ is then:
\begin{equation}
\theta(k,t) = 2 k \sum_{n = 1} C_n(k)~\frac{(-1)^{n-1}}{n!}~t^n
\label{PE} \end{equation}
where the first order in time is $2 k t$, because the first
adsorption attempt on an empty lattice is always accepted and
occupies $k$ sites in the two possible orientations,
and where the coefficients $C_n(k)$ are given by implicit overlap
integrals: \begin{equation}
C_n(k) = \prod_{i=2}^n ( 1 - \prod_{j=1}^{i-1} ( 1 - K_{i~j}))
\quad , n \geq 2 \quad.
  \label{C_n}
\end{equation}
The set of integration
 variables defining a deposition is denoted by $i$ , and  $-K_{i~j}$
is a hard Mayer function ($K_{i~j}=1$ only if $i$ and $j$ are
overlapping objects, $0$ otherwise).

The inclusion-exclusion sequence generated by Eqs.(\ref{PE}-
\ref{C_n}) has already been used for square deposition in the
continuum \cite{DWJ,BHM}.
Similar techniques
appear in hard sphere modelization of simple liquids at equilibrium
\cite{DWJ,TST}.

Next, one generates the n-th order diagrams by the full expansion of
$C_n(k)$. All monomials of this polynomial are
connected labeled graphs which are regrouped in classes
$\Gamma_{n,i}$ (unlabeled graphs) of the same topology which
thus appear with combinatorial weights as can be easily seen in Fig.
5
which shows the first terms of the graphical perturbative expansion.

Let us stress  the generality of the above perturbative approach which
is valid for any standard RSA process.
   The details of the process affect only the graph integrals
         $I(\Gamma_{n,i})$.

 In order to compute $I(\Gamma_{n,i})$, one has to do the summation
over
 $\{ x_l \}$, the n vertices of the graph $\Gamma_{n,i}$. Here
the vertices, $\{ x_1, x_2, \ldots , x_n \}$, are to be understood as
the position of the starting point and the orientation of each of the
$n$ $k$-mers.

Finally, the graph contribution is  \begin{equation}
 I(\Gamma_{n,i}) = \sum_{\{ x \}}~\prod_p K(x_{l_p},x_{m_p})
 \label{graph} \end{equation}
where the product $\prod_p$ is done on the links of $\Gamma_{n,i}$
and the
 sum is over all the degrees of freedom for the $k$-mer variable
$x_l$, except  one which can be frozen at the origin by transational
invariance.

Aligned object deposition is of practical interest here because, due
to the factorization property of hard Mayer functions
\cite{deRocco},
 it allows us to reduce the
problem of two-dimensional integration to one-dimensional one, once
the orientations of the $k$-mers are fixed.
Such a factorization has already been used in the RSA of aligned
hypercubes to determine the time-series \cite{DWJ,BHM} and proved to
be essential in the study of the Pal\'asti conjecture \cite{BHM}.
In practice, we have analytically
calculated all the $2^n$ projected one-dimensional graph integrals
appearing at order n. Sums of products of two 1-d integrals are then
performed to algebraically obtain the 2-d graph integrals for
segment deposition.

We have thus stored all the
one dimensional graph integrals $I_{n,i}$ and
weights necessary to compute $C_n(k)$ analytically in $k$, up to the
seventh order in $t$. The analytic expression of graph integrals has
no physical interest and we give in Table III the coefficients
needed to reconstruct
to  $7^{th}$ order   the time series expansion of which the
first terms are
\begin{equation}
\theta(k,t) =  2 k~( \quad t~ - (- 1 + 2 k   + k^2
)~\frac{t^2}{2}~
+ ( 1 - 5 k   + k^2 + 5 k^3 + 2 k^4 )~\frac{t^3}{6} \quad   +
\ldots	  )
\label{PEfirst}
\end{equation}

We give, as a first attempt at resummation,  the predictions for
the
 dimer jamming coverage $\theta_{\infty}$ from one of the more
  stable methods for computing the infinite time limit of
  $\theta(2,t)$, knowing
the exponential behaviour of the coverage at large times on the
lattice.

We first invert the PE, Eq.(\ref{PE}), into a power
series of $\theta$, to calculate $e^{t}$ in terms of $\theta$.
The poles of the Pad\'e approximants formed from this series in
$\theta$ give  jamming values $\theta_\infty$, which are reached
with
the expected exponential behaviour
$\theta(t) \simeq \theta_{\infty} - A e^{- t} $, the coefficient
$A$ being given by the residue. The resulting Pad\'e tables of
$\theta_{\infty}$
of the coefficients $A$ can be found in Table IV,
where we notice very good agreement between
simulations  and time-series estimates of the jamming
coverage. In fact, this series for deposition of small
size objects
 can be numerically performed at larger orders than the generic
case of $k$-mer deposition, and allows us to observe a wider Pad\'e
table.

The quality of the results, linked to the dispersion of the Pad\'e
table, decreases with the length of the $k$-mer. For example, the
same method applied to the $8^{th}$ order trimer series is still
predictive, giving $\theta_{\infty} = 0.842(2)	 $, and $A= 0.135(5)
$ in agreement with both the result of Evans and Nord \cite{EN},
i.e. 0.8465 obtained by hierarchy truncation which exploids empty
site shielding, and the  Monte-Carlo simulation result of Nord
\cite{Nord}, 0.8465(2).
For $k \geq 4$, numerous instabilites forbid reliable evaluations.

These instabilities are reminiscent of the problems encountered in
summing up the RSA series of ($k \times k$) squares deposition	on a
square lattice as $k$ increases: the effective behaviour of the
truncated power series
seems to change, as $k$ increases, from $e^{-t}$ to $ \ln{t}/t$,
which prevents the use of a simple and stable  extrapolation
procedure
 in the intermediate $k$ range.
The situation is  even worse for the deposition of long $k$-mers
        because
the usual scaling argument does not  hold, in such a way that
the infinite $k$ limit of the time
series expansion of the coverage cannot be taken order by order.
This can then be seen in the behaviour of the series for
large $k$ \begin{equation}
\theta(k,t) \simeq \frac{1}{k} \sum_{n\geq 1} N_n \quad
(k^2 t )^n ,\qquad k \gg 1 ,
\label{wrongscaling}
\end{equation}
which implies either $\theta(k,\infty) = O(1/k)$ or the divergence of
the sum.

The next section will report on a large $k$  re-summation
procedure and  compare with the Monte-Carlo results of
section 2.

\subsection{Summation of the perturbative  expansion in the large $k$
regime} ~

In this section we assume that the long $k$-mer limit of the jamming
coverage does not vanish, as shown by our simulation results.

Let us first discuss the connection between adsorption of line
segments by a one-dimensional lattice and adsorption by a two-
dimensional one, as  already suggested in section 2.
Line segment deposition on a square lattice contains obviously
line segment deposition on each of the one dimensional sub-lattices
and in the correlation	between these two competing (horizontal and
vertical) adsorptions resides  the difficulty of the study.
Owing to the flux definition  used in Eq.(\ref{PE}),
the one dimensional coverage $\theta_1(k,t)$ appears
explicitly in both lattice directions as one special
configuration among all the others
because one has to sum over all possible relative orientations of the
segments.
 This is perturbatively observed on the time-series of the coverage in
which
 $2 \theta_1(k,t)$ sums up all the terms in $k$ and $k^2$.
Moreover we can rewrite Eq.(\ref{wrongscaling}) as
\begin{equation}
\theta(k,t) \simeq 2 k t ~\sum_{n\geq 0} \tilde{N}_n \quad (k^2 t )^n
,\quad k \gg 1 ,
\label{goodscaling}
\end{equation}
which then shows two kind of ``scaling'' variables: $ k t$
typical of a
one-dimensional RSA of $k$-mers and a second one $k^2 t$.

 Therefore we define a	``correlation'' function by
\begin{equation}
 \Gamma (k,t) = \frac{\theta(k,t)}{2 \theta_1(k,t)}   \qquad .
\label{gamma}
\end{equation}

Using the results from the Monte-Carlo simulation  of Table I and
from the series Eq.(\ref{PEfirst}),
inserting the exactly known  one dimensional coverage
$\theta_1(k,t)$, we can give the following properties of
$\Gamma(k,t)$, which illustrate its smooth behaviour:
\begin{itemize}
\item $\Gamma(k,0) = 1	\qquad \forall k$
\item $\Gamma(k,\infty)$, which is related to the jamming
      coverage, varies slowly with  $k$ \\
 (e.g. $\Gamma(k,\infty) = 1/2, .524, .514, .504, \dots,  .444$
	for $k=1,2,3,4, \dots, \infty$).

\item We can define a scaling variable $u = k^2 t$ in such a way that
the
 coefficients of the power series in $u$ of $\Gamma(k,u)$ are slowly
varing
 polynomials in $1/k$, as it can be seen from the first terms of
    $\Gamma(k,u)$ $$ \Gamma(k,u)=1  -\frac{u}{2}
+  (1+\frac{1}{k}-\frac{5}{4 k^2}) \frac{u^2}{3} +\ldots \qquad .$$
\end{itemize}

Collecting the sub-series in u of a given power of $1/k$ in
$\Gamma(k,u)$ we then define
\begin{equation}
\Gamma(k,u)=
\Gamma(\infty,u)~~(1~+\frac{G_1(u)}{k}~+\frac{G_2(u)}{k^2}+\ldots)
\quad , \label{gammaku}
\end{equation}
which has to be understood as an asymtotic expansion of $\Gamma(k,u)$
when $k$ goes to infinity, as long as the various series in $u$
appearing in Eq.(\ref{gammaku}) can be resummed, and in particular
\begin{equation}
 \Gamma (\infty,u) =   1  -\frac{1}{2} u      + \frac{1}{3}  u^2
- \frac{23}{108}  u^3 + \frac{283}{2160} u^4 - \frac{50593}{648000}
          u^5
 + \frac{264017}{5832000}  u^6	+O(u^7) \quad .
\label{gammau}
\end{equation}

In the following  we shall work out a resummation procedure for
$ \Gamma (\infty,u)$  together with an evaluation of the subleading
terms
in $1/k$, and we find sensible results.

A quick glance at the perturbative expansion of $\Gamma(\infty,u)$
given by Eq.(\ref{gammau}) reveals its striking similarity  with
the expansion of $\ln(1+u)/u$, starting with an identity of the first
three terms. Therefore we write
\begin{equation}
\Gamma(\infty,u) =\frac{ \ln ( 1 + \Phi(u) )}{\Phi(u)}
\label{gammaphi}
\end{equation}
in which the perturbative expansion of $\Phi(u)$ can be easily
obtained
from Eqs.(\ref{gammau}-\ref{gammaphi}).

The final task is to find the limit at $u=\infty$ of $\Phi$.
We have used a standard method for such an extrapolation, namely a
mapping $v(u)$ of the $u$ variable followed by a Pad\'e analysis of
the $v$ series. In practice we have searched for intersections of
Pad\'e approximants by varying the parameter $\alpha$ entering the
definition of the mapping
$$ v(u) = \frac{1-e^{-\alpha u}}{\alpha}$$
after the approximants have been calculated at the point
$v=1/\alpha$.
{}From these intersections leading to
the evaluation of $\Phi(\infty)$ we finally obtain
$\theta(\infty,\infty)$.

Fig. 5	shows the Pad\'e intersections in the plane $(\theta,
\alpha)$.
The multiple intersections (of the $5^{th}$ and $6^{th}$ order)
group into four nearby classes in which
we select those containing the approximants of the highest order
by $\theta=0.658$ and $\theta=0.670$  and thus	 we deduce our
result
 for the coverage of a square lattice by infinitely long line-
segments $$\theta(\infty,\infty) = 0.664(6)~~,$$
in good agreement with the Monte-Carlo value, $\theta=0.660(2)$,
obtained in section 2.

Lets us now briefly describe the time series computation of the sub-
leading coefficients $A_1$ and $A_2$ which appear in the asymptotic
expansion of the jamming coverage
$$\theta(k,\infty)=
\theta(\infty,\infty)+\frac{A_1}{k}+\frac{A_2}{k^2}+\ldots \qquad
.$$

We have applied the same mapping and Pad\'e analysis as above  on
the sub-dominant series $G_1(u)$ and $G_2(u)$ defined by
Eq.(\ref{gammaku}). Among  the	distinct solutions that we have
obtained, namely \\
$( G_1(\infty), \alpha ) = \{ (0.956,0.887),(0.713,1.165),
(1.889,0.333)\}$ and \\$( G_2(\infty), \alpha )= \{ (-1.377,0.694), (-
0.791,1.368)\}$,\\
      the prefered solution, which corresponds to a larger
       number of
intersecting central approximants, has been quoted
first.
 These values of $G_1(\infty)$ and $G_2(\infty)$ and
the known asmptotic behaviour of $\theta_1(k,\infty)$
\cite{MacKenzie} allows us to finally give our prefered
estimates for $A_1$ and $A_2$ $$A_1 = 0.827  ,\qquad  A_2= -
0.699$$
together with the global range we have obtained from this method
$$ 0.6 \leq A_1 \leq 1.5 ,\qquad  -0.8 \leq A_2 \leq -
0.1\mbox{\quad .}$$ This result compares well to the best fit of
our Monte-Carlo data from Eq.(\ref{MCfit}), except for $A_2$ ,
found too small.
As a consistency check of the $1/k$ expansion we have also
resummed $\Gamma(k,u)$ for finite  values  of $k$ using the same
method as for $\Gamma(\infty,u)$. We find essentially the same
kind of results,
which only differ by  few $\%$ from our asymptotic calculation
which is $$\theta(k,\infty)= 0.664 + \frac{0.827}{k} -
\frac{0.699}{k^2} \qquad ,$$ and gives an unexpected precision of
the coverage at all $k$ values. Actually it deviates from the
simulation data of Table I at most by $2\%$, reached at $k=4$, in
the whole $k$ range.

In conclusion of this section we have shown that the perturbative
expansion summation of long $k$-mer coverage can be brought into
full agreement with the Monte-Carlo simulations.
\section{Conclusion}
{}~\\
We have determined the saturation coverage of randomly adsorbed
segments on a square lattice as a function of the size of the
segments, by two independent methods : a numerical simulation and a
time series resummation method. Its behaviour for large segments has
been obtained and both methods give comparable results, both for the
asymptotic value itself and for the approach to this limit.
 Indications that the large segment
deposition process is driven by a one-dimensional mechanism have been
seen in both cases, on the one hand through linear fluctutations in
the simulation and on the other by   the particular role played by
the
one-dimensional coverage time-series that we used in the resummation
        of the
two-dimensional one.

Actually, a simple argument may explain this fact.
 If one assume that a jammed
configuration is translationally invariant (in average) and invariant
versus the exchange of the $X$ and $Y$ axis, it is sufficient to
determine the coverage of {\em a single line} of the system to get
the total coverage. On a line, the occupied sites are distributed
among segments and points, which result from the intersection of the
line with transversely deposited segments. This suggests that a mean-
field approach can successfully describe this model. If this were
indeed true, it would be possible to derive some simple
approximations for
the correlations functions. These are particularly useful for
characterizing the
jammed configurations, where some local order is clearly apparent.
This order is caused
by the tendency
of the last deposited segment to align with previously deposited
ones.
We plan to investigate this aspects of the model in the future.
\vspace{2cm}\\

\newcommand{\rev}[4]{{\em #1} {\bf #2}, #3 (#4)}
\newcommand{\pr}[3]{{\em Phys. Rev.} {\bf #1}, #2 (#3)}

 \newpage

\section*{Table captions}
\begin{itemize}
\item[{\bf Table I}] - The jamming limit for the deposition of $k$-mers on
lattice of size $L$, for various segment lengths $k$.

\item[{\bf Table II}] - The standard deviation $\sigma_\theta$ of the
coverage for different fixed values of $L/k$ as a function of the
lattice size.

\item[{\bf Table III }] -
This  table gives the coefficients $\alpha_n^p$
allowing us to
reconstruct the $7^{th}$-order time-series expansion of $
\theta(k,t)$ , the lattice coverage by segments of $k$ sites,
by
$\theta(k,t) = 2 k \sum_{n=1} \frac{(-1)^{n-1}}{n!}~ C_n~ t^n$ ,
in which $ C_n =  \sum_{p=0}^{2 n - 2} ~ \alpha_n^p~ k^p$ . For
small $k$-mers,
more orders can easily be computed: $k=2, C_8=73035123,
C_9=1663498315 $, and for $k=3, C_8=20554179608 $.

\item[{\bf Table IV }] -
Pad\'e table of the dimer jamming coverage obtained
through
$ \exp{t} = Pade[ N(\theta) / D(\theta) ]$ where the
denominator $D$ (numerator $N$) degree varies horizontally
(vertically), respectively.
For each entry of the first  table the coefficient A of the
approach to the jamming limit given by
$\theta \simeq \theta_{\infty} - A e^{-t}$
 has been computed through
the residue of the corresponding pole, and can be found in the
second table. For comparison we recall the simulation estimates of
0.9069(2)
from Nord \cite{Nord} and of 0.9068(1) from Table I.\\
 \end{itemize}

 \newpage
{}~\\
\begin{center}
{\tiny
\begin{tabular}{lllllllll}
\hline
\hline
&&&&&&&&\\
$\;\; L$ & 128 & 256 & 512 & 1024 & 1536 & 2048 & 3072 &
4096 \\
$k$ &&&&&&&&\\
 \hline
&&&&&&&&\\
 2  & 0.9066(2) & 0.9068(1) & &&&&&\\
 4  & 0.8109(3) & 0.8106(2) & 0.8106(1) &&&&&\\
 8  & 0.7487(6) & 0.7484(3) & 0.7477(1) & 0.7477(1)
&&&&\\
 12 & 0.7233(8) & 0.7239(4) & 0.7239(2) & &&&&\\
 16 & 0.7107(13) & 0.7111(6) & 0.7106(2) &
0.7110(1) &&&&\\
 24 & 0.6956(6) & 0.6970(7) & 0.6968(4) & 0.6967(2)
&&&&\\
 32 & 0.6867(23) & 0.6894(13) & 0.6890(7) &
0.6893(4) &&&&\\
 48 & & 0.6812(20) & 0.6814(9) & 0.6809(5)  &&&&\\
 64 & & 0.6721(25) & 0.6769(13) & 0.6765(6)  &&&&\\
 96 & 0.6709(38) & 0.6734(20) & 0.6731(6) & 0.6714(5) &&&\\
 128& & & 0.6697(24) & 0.6692(13) & & 0.6682(6)  &&\\
 192& & & & & 0.6656(13) & & 0.6655(7) & \\
 256& & & & 0.6608(27) & & 0.6641(10) & & 0.6637(6)\\
 384& & & & & & & 0.6632(13)& 0.6634(6)\\
 512& & & &  & &  & & 0.6628(9) \\
&&&&&&&&\\
\hline
\hline
\end{tabular}
}
\vfill
{\bf Table I}
\end{center}

 \newpage
{}~\\
\begin{center}
\begin{tabular}{llllll}
\hline
\hline
&&&&&\\
$ \frac{L}{k}\;\setminus\; L$ & 64 & 128 & 256 & 512 & 1024 \\
&&&&&\\
 \hline
&&&&&\\
32 & 0.0038(2) & 0.0034(2) & 0.0030(2) & 0.0024(2) & 0.0031(3)\\
16 & 0.0070(3) & 0.0060(3) & 0.0060(3) & 0.0059(4) &
0.0061(4)\\ 8 & 0.0125(4) & 0.0122(4) & 0.0119(7) &
0.0117(8) & 0.0129(7)\\
4 & 0.0247(9) & 0.0251(13) & 0.0235(9) & 0.0265(10) & 0.0267(16)\\
&&&&&\\
\hline
\hline
\end{tabular}\vfill
{\bf Table II}
\end{center}

\newpage
\begin{center}
\begin{tabular}{lrrrrrrr}
\hline
\hline
&&&&&&&\\
\multicolumn{1}{c}{ }    &
\multicolumn{1}{c}{1}    &
\multicolumn{1}{c}{2}    &
\multicolumn{1}{c}{3}    &
\multicolumn{1}{c}{4}    &
\multicolumn{1}{c}{5}    &
\multicolumn{1}{c}{6}    &
\multicolumn{1}{c}{7}      \\
&&&&&&&\\
\hline
&&&&&&&\\
$k^{ 0}$ & 1& -1&  1& -  9& 18&  -900&
8100	    \\
$k^{ 1}$ &  &  2& -5& 84&   -276& 20940&  -
269460	    \\
$k^{ 2}$ &  &  1&  1& - 95&    645&    -79802&
1459620     \\
$k^{ 3}$ &  & &  5& -141&    343& 46090& -
3561354     \\
$k^{ 4}$ &  & &  2& 22&   -323&    -88615&
11703689    \\
$k^{ 5}$ &  & &   &  165&  -2288&    648530& -
33506220    \\
$k^{ 6}$ &  & &   & 46&    529&   -365541&
16916830    \\
$k^{ 7}$ &  & &   &   &   1357&   -745840&
59786163    \\
$k^{ 8}$ &  & &   &   &    283&    235865& -
43362258    \\
$k^{ 9}$ &  & &   &   &   &    307480& -
40950720    \\
$k^{10}$ &  & &   &   &   & 50593&
16739500     \\
$k^{11}$ &  & &   &   &   &      &
13706391     \\
$k^{12}$ &  & &   &   &   &      &
1848119     \\
&&&&&&&\\
\hline
\hline
\end{tabular}
\vfill
{\bf Table III }
\end{center}

 \newpage
 ~\\
\begin{center}
\begin{tabular}{ccccccccc}
\hline
\hline
$[ N , D ]$ & 1  & 2     &       3      &    4  &  5
   &       6       &      7    &    8   \\
\hline
1  &  1       &   .94007   &   .92212   & .91478    &
.91127
   &  .90943       &   .90841   &   .90781   \\
2  &  .94117       &   .91535   &   .91011   & .90826    &
.90750
   &  .90716       &   .90699   &    \\
3  &  .92307       &   .91015   &   .90781   & .90718    &
.90696
   &  .90688       &     &    \\
4  &  .91549       &   .90831   &   .90718   & .90692    &
.90686
   &        &     &    \\
5  &  .91179       &   .90753   &   .90696   & .90686    &
   &        &     &    \\
6  &  .90980       &   .90718   &   .90688   &       &
   &        &     &    \\
7  &  .90867       &   .90701   &    &       &
   &        &     &    \\
8  &  .90800       &     &    &       &
   &        &     &    \\
\hline
\hline
\end{tabular}

\vspace{1cm}
 ~\\
\begin{tabular}{ccccccccc}
 \hline
 \hline
$[ N , D ]$ & 1  & 2     &       3      &    4  &  5
   &       6       &      7    &    8   \\
\hline
1  & .25         &  .20744    &  .19183   & .18416    &
.17986
   & .17726       &  .17562    &  .17455   \\
2  & .20843       &  .18425    &  .17788   & .17513    &
.17379
   & .17310       &  .17273    &    \\
3  & .19285       &  .17794    &  .17433   & .17312    &
.17263
   & .17244       &     &    \\
4  & .18506       &  .17521    &  .17313   & .17254    &
.17238
   &        &     &    \\
5  & .18061       &  .17386    &  .17264   & .17238    &
   &        &     &    \\
6  & .17788       &  .17315    &  .17244   &       &
   &        &     &    \\
7  & .17611       &  .17276    &    &       &
   &        &     &    \\
8  & .17494       &     &    &       &
   &        &     &    \\
\hline
\hline
\end{tabular}
\vfill

{\bf Table IV }
\end{center}

\newpage
\section*{Figure captions}
\begin{itemize}
\item[{\bf Figure 1 :}] The jamming limit for the deposition of $k$-
mers on lattice of
size $L=128$ (open circles) and $L=256$ (filled circles) as a function
of $k$. The solid line corresponds to the best fit of the
$L\rightarrow\infty$ values. \item[{\bf Figure 2 :}] The standard
deviation $\sigma_\theta$ as a function of the
segment size $k$ for the $L=256$ (open circles) and $L=512$ (filled
circles) lattice sizes. The solid line is the best fit of the $L=256$
data by a linear function.
\item[{\bf Figure 3 :}] The estimated $L\rightarrow\infty$ jamming
coverage as a function of the segment size $k$ and the fit of
these data.
\item[{\bf Figure 4 :}]
Graphical expansion of $C_n$ for  1, 2, 3 and 4 points. \\
The reduction of the
the number of graphs operated by the topological identification is
especially efficient at large orders not depicted on this figure,
 e.g. at $7^{th}$ order, $\simeq 6 ~~ 10^5$
labeled graphs are regrouped into $\simeq 10^3$ unlabeled ones.
\item[{\bf Figure 5 :}]
Pad\'e approximants plot of the $5^{th}$ and $6^{th}$ order versus the
variational parameter $\alpha$.\\
Intersections of Pad\'e approximants of the $6^{th}$ order are shown by
full circles, whereas open circles are located at the other
intersections.
\end{itemize}

\end{document}